\documentclass[a4paper,11pt]{article}

\usepackage[utf8]{inputenc}
\usepackage{geometry}
\usepackage{fancyhdr}
\usepackage{setspace}
\usepackage{hyperref}
\usepackage{graphicx}
\usepackage{float}
\usepackage{multirow}

\usepackage[style=apa,backend=biber]{biblatex}
\begin{document}

\begin{center}
    {\Large{Disappearing repositories -- taking an infrastructure perspective on the long-term availability of research data}}
\end{center}

\vspace{0.5cm}

\begin{itemize}
\centering
    \item[] Dorothea Strecker\textsuperscript{1} (corresponding author)\\dorothea.strecker@hu-berlin.de\\
    \href{https://orcid.org/0000-0002-9754-3807}{0000-0002-9754-3807}
    \item[] Heinz Pampel\textsuperscript{1,2}\\
    \href{https://orcid.org/0000-0003-3334-2771}{0000-0003-3334-2771}
    \item[] Rouven Schabinger\textsuperscript{3}\\
    \href{https://orcid.org/0000-0002-0249-7917}{0000-0002-0249-7917}
    \item[] Nina Leonie Weisweiler\textsuperscript{2}\\
    \href{https://orcid.org/0000-0001-6967-9443}{0000-0001-6967-9443}
\end{itemize}

\begin{center}
    $^1$ Humboldt-Universität zu Berlin, Berlin School of Library and Information Science\\
    $^2$Helmholtz Association, Helmholtz Open Science Office\\
    $^3$Swiss Library Service Platform (SLSP)
\end{center}

\vspace{1cm}

\section*{Abstract}
Currently, there is limited research investigating the phenomenon of research data repositories being shut down, and the impact this has on the long-term availability of data. This paper takes an infrastructure perspective on the preservation of research data by using a registry to identify 191 research data repositories that have been closed and presenting information on the shutdown process.
\\
The results show that 6.2 \% of research data repositories indexed in the registry were shut down. The risks resulting in repository shutdown are varied. The median age of a repository when shutting down is 12 years. Strategies to prevent data loss at the infrastructure level are pursued to varying extent. 44 \% of the repositories in the sample migrated data to another repository, and 12 \% maintain limited access to their data collection. However, both strategies are not permanent solutions.
\\
Finally, the general lack of information on repository shutdown events as well as the effect on the findability of data and the permanence of the scholarly record are discussed.

\section*{Keywords}
research data ; research data repository ; infrastructure maintenance ; scholarly record

%\maketitle
\thispagestyle{empty}

\newpage
\onehalfspacing
\pagenumbering{arabic}

\section{Introduction}
With the number of published research data steadily increasing \parencite{benjelloun_google_2020}, the long-term preservation of datasets is gaining importance, especially if research data are to be regarded as self-contained components of the scholarly record \parencite{manghi_new_2021}. For this idea and data citation to succeed, continuous access to datasets is required, since in order for datasets to become citable units, they must be permanently available \parencite{buneman_data_2021}.
Concerns about perpetual access to digital scholarly texts have resulted in the establishment of a distributed network of preservation services that is maintained jointly by various stakeholders \parencite{mering_preserving_2015}. However, the adoption of these preservation services is slow compared to the growth in the number of academic journals, and some journals have been shut down and disappeared \parencite{laakso_open_2021}. Research data might be even more vulnerable, as the burden of long-term preservation rests predominantly on dedicated repositories -- preservation systems comparable to those for scholarly texts currently are not widely spread and can be difficult to realize \parencite{kiefer_digital_2015}.
\\
Long-term preservation of research data requires not only continuous care of datasets, but also of the repositories that hold them \parencite{eschenfelder_organizational_2017}. The TRUST Principles, a set of guiding principles for research data repositories formulated by a multi-stakeholder group, ask repositories to “ensure uninterrupted access to [their] valuable data holdings for current and future user communities”\parencite[3]{lin_trust_2020}. To meet these expectations, repository operators need to find solutions for infrastructure maintenance, governance, securing of funding, and continuity planning. The long-term operation of research data repositories presents a challenge, and sometimes, for varying reasons and despite best efforts, research data repositories are shut down.
\\
Beyond case studies of individual repositories or selected disciplines, there is currently little research investigating the phenomenon of research data repositories being shut down and the impact this has on the long-term availability of data \parencite{boyd_understanding_2021}. To address this issue, this paper identifies research data repositories of all types and from all disciplines that have been closed and presents information on the shutdown process.
\\
Based on metadata from a registry of research data repositories and information collected from repository websites, this paper takes an infrastructure perspective on the long-term preservation of research data. The following questions are addressed:
\begin{itemize}
    \item How common is the phenomenon of research data repositories being shut down?
    \item What are the risks that lead to repositories being shut down?
    \item What measures are taken to prevent data loss when repositories are shut down?
\end{itemize}
The paper concludes with reflections on how repositories and registries can contribute to better documentation of repository shutdown and data migration.

\section{Background}

\subsection{Research data repositories}
Infrastructures are ``pervasive enabling resources in network form'' \parencite[98]{bowker_toward_2010}. They permeate specific areas of life and support certain practices \parencite{laak_lifelines_2023}. In the area of knowledge work, information infrastructures form ``robust networks of people, artifacts, and institutions that generate, share, and maintain specific knowledge about the human and natural worlds.'' \parencite[17]{edwards_vast_2013} Research data repositories are specialized information infrastructures that focus on the curation, preservation, and dissemination of research data \parencite{boyd_understanding_2021,edwards_vast_2013, johnston_how_2018}. They facilitate data journeys -- the movements of data from the sites of production to the sites of (re-)use -- and are central components for realizing the vision of habitual data publication, a fundamental Open Science practice \parencite{austin_key_2017,bates_data_2016}.
\\
Although studies on researchers’ attitudes towards publishing their data in repositories are still inconclusive \parencite{thoegersen_researcher_2021}, there is evidence that the use of research data repositories to make data available has increased in recent years \parencite{jiao_data_2022, khan_data_2023}.
Research data repositories shape and are shaped by the communities they serve. They can become sites of scientific collaboration by enabling researchers to gather around datasets they jointly use \parencite{costa_research_2014, lafia_subdivisions_2022,qin_structural_2022}, and might have to adapt services if the designated community it serves shifts over time \parencite{donaldson_data_2020}. At a global level, the landscape of research data repositories has continuously evolved. Driven by the ideal of promoting Open Access to scholarly publications, the number of repositories increased significantly between 2005 and 2012 \parencite{pinfield_open-access_2014}, initially focusing on text publications. Later, infrastructures specialized in the management of research data gradually emerged, forming a global network of heterogeneous research data repositories \parencite{kindling_landscape_2017}. In 2023, the global registry re3data lists more than 3.000 research data repositories.
Different types of research data repositories have evolved to serve specific needs, for example institutional repositories that support members of research and higher education organizations \parencite{arlitsch_why_2018}, discipline-specific repositories that address researchers from a particular research area \parencite{banzi_evaluation_2019}, or generalist repositories, which allow the storage of data irrespective of discipline \parencite{stall_shelley_generalist_2020}.
\\
Overall, the publication of research data appears to be characterized by concentration tendencies. For example, a recent study of a data discovery service found that only 20 repositories accounted for almost 80 \% of the total datasets indexed \parencite{benjelloun_google_2020}. Most research data repositories are also operated by institutions located in Europe and North America \parencite{kindling_landscape_2017}, and in the last few years, European research data repositories have grown substantially in size \parencite{dans_european_2022}. The number of repositories in Africa, Asia, and BRICS countries has increased in recent years, but operators of these repositories sometimes find it challenging to fully realize their visions, for example if funding or institutional support are lacking \parencite{academy_of_science_of_south_africa_assaf_african_2019, cho_study_2019, misgar_study_2020, nishikawa_how_2020}.

\subsection{Long-term operation of research data repositories}

\subsubsection{Time scales of research data repositories}
Time is a challenge for all types of infrastructures, because they operate at two different time scales at once: They must both be usable now and remain usable in the long term. As Karsati et al. put it, “an infrastructure occurs when here-and-now practices are afforded by temporally extended technology” \parencite[400]{karasti_infrastructure_2010}. Both time scales can threaten the existence of a research data repository – it might be shut down if it is unable to serve current needs and practices, or if it can’t offer reliable services over a long period of time. Repository operators are aware of these time scales, as many have expressed concerns both about the long-term maintenance of their repository and the ability to develop new functionalities \parencite{khan_are_2021}.
\\
A research data repository also has to bridge another time gap: The varying life spans of its technical components and its data collection. Likely, the life span of a data collection is longer compared to the technical components required for its preservation: ``data collections accrue slowly and steadily, yet software and hardware can change relatively rapidly and are beyond the control of collection staff responsible for data collection.'' \parencite[4]{thomer_patchwork_2023} Therefore, in order to preserve data collections, the research data repositories storing them must also be maintained.

\subsubsection{Risks to the long-term operation of research data repositories}
Planning for the long-term preservation of research data is challenging, because various factors can put both the data and the repository that holds them at risk \parencite{thomer_supporting_2018}. Research data are at risk of being lost if the research data repository is threatened, for example if it is facing loss of funding \parencite{mayernik_risk_2020}. In interviews, developers and auditors of repository standards as well as repository staff identified five potential sources of risk: finance, legal, organizational governance, repository processes, and technical infrastructure \parencite{frank_risk_2022}.
Barateiro et al. developed a comprehensive typology of risks to digital long-term preservation systems, such as research data repositories \parencite[8ff]{barateiro_designing_2010}. The typology lists vulnerabilities (``weaknesses [...] in the environment'') and threats (``events that affect normal behaviour'') that can adversely impact components of these systems (see Table \ref{tab:risk-typology}). Vulnerabilities can be introduced to preservation systems either by software (process), by characteristics of the information objects being preserved (data), or by the infrastructure (infrastructure). Preservation systems can be threatened by non-deliberate actions (disasters), deliberate actions (attacks), managerial decisions (management), or changes in laws (legislation).
\begin{table}[H]
\centering
\small
\begin{tabular}{|p{2.7cm}|p{5.5cm}|p{5cm}|}
\hline
\multirow{8}{*}{vulnerabilities} & \multirow{2}{*}{\begin{tabular}[c]{@{}l@{}}process\end{tabular}} & software faults          \\ \cline{3-3} 
                                 &                                                                                                                                                   & software obsolescence    \\ \cline{2-3} 
                                 & \multirow{2}{*}{\begin{tabular}[c]{@{}l@{}}data\end{tabular}}               & media faults             \\ \cline{3-3} 
                                 &                                                                                                                                                   & media obsolescence       \\ \cline{2-3} 
                                 & \multirow{4}{*}{\begin{tabular}[c]{@{}l@{}}infrastructure\end{tabular}} & hardware faults          \\ \cline{3-3} 
                                 &                                                                                                                                                   & hardware obsolescence    \\ \cline{3-3} 
                                 &                                                                                                                                                   & communication faults     \\ \cline{3-3} 
                                 &                                                                                                                                                   & network service failures \\ \hline
\multirow{8}{*}{threats}         & \multirow{2}{*}{\begin{tabular}[c]{@{}l@{}}disasters\end{tabular}}                                        & natural disasters        \\ \cline{3-3} 
                                 &                                                                                                                                                   & human operational errors \\ \cline{2-3} 
                                 & \multirow{2}{*}{\begin{tabular}[c]{@{}l@{}}attacks\end{tabular}}                                              & internal attacks         \\ \cline{3-3} 
                                 &                                                                                                                                                   & external attacks         \\ \cline{2-3} 
                                 & \multirow{2}{*}{\begin{tabular}[c]{@{}l@{}}management\end{tabular}}                                    & economic failures        \\ \cline{3-3} 
                                 &                                                                                                                                                   & organizational failures  \\ \cline{2-3} 
                                 & \multirow{2}{*}{\begin{tabular}[c]{@{}l@{}}legislation\end{tabular}}                                  & legislative changes      \\ \cline{3-3} 
                                 &                                                                                                                                                   & legal requirements       \\ \hline
\end{tabular}
\caption{Typology of risks to long-term preservation systems, \cite[9]{barateiro_designing_2010}}
\label{tab:risk-typology}
\end{table}
\noindent
Risks to the long-term operation of research data repositories have long been discussed in the literature, in particular the risk of economic failure \parencite{chowdhury_sustainability_2013}. Concerns about funding cuts that threaten databases in the life sciences have been raised for years \parencite{baker_databases_2012, merali_databases_2005}. If funding for databases is discontinued, entire research communities in the life sciences can be impacted, as the cases of the Biological Magnetic Resonance Bank (BMRB) \parencite{nature_structural__molecular_biology_support_2012}, EcoCyc \parencite{sri_international_ecocyc_2014}, and Online Mendelian Inheritance in Man (OMIM) \parencite{kaiser_funding_2016} exemplify. In these cases, researchers who used the repositories publicly called to save the databases, as they were considered essential resources. These initiatives were successful, since the databases were still operational at the time this paper was written. Economic sustainability remains a major concern today \parencite{ficarra_scoping_2020}, although the funding of research data repositories has become more reliable in part \parencite{burns_institutional_2013}. Revenue streams of research data repositories can change over time \parencite{eschenfelder_financial_2022}, but still depend predominantly on publicly funded organizations \parencite{imker_who_2020}.
\\
Repositories also try to reduce technical risks (process and infrastructure vulnerabilities) \parencite{eschenfelder_organizational_2017}. Generally, infrastructures such as research data repositories can be adapted and reconfigured, but the “installed base”, components they build on, might limit this flexibility and therefore pose a risk \parencite{hirsch_sedimentary_2022}. These limits can contribute to tensions between building something new and maintenance work \parencite{ribes_tensions_2007}.
\\
Ensuring the sustainability of a research data repository is a continuous process \parencite{eschenfelder_organizational_2017}. Despite this fact and the importance of infrastructure maintenance for the long-term preservation of data, it is not explicitly included in widely used models of digital curation, for example lifecycle models or the OAIS reference model \parencite{thomer_three_2020}.

\subsection{Research data repositories being shut down}
One characteristic of infrastructures in general is that they are often taken for granted and only noticed once they break down \parencite[113]{star_steps_1996}. This is also true for information infrastructures such as research data repositories, but the point when they ``shut down'' is not always clear cut; for example, the information infrastructure can be closed entirely, scaled down, or disassembled into components that are repurposed later \parencite{steinhardt_breaking_2016}.
\\
Despite careful planning, a research data repository might be closed eventually \parencite{dean_shutting_2016}. Currently, there is little evidence of how often this phenomenon occurs. A study of biological databases found that after a period of 18 years, 75 \% were either closed entirely or the content was no longer updated \parencite{attwood_longevity_2015}, but this high rate of shutdowns might not translate to other repository types or disciplines. In another case, a consortial repository was closed after former consortium members gradually reorganized research data management and moved their data to institutional repositories \parencite{dean_shutting_2016}. A recent study evaluated link rot in four registries of scholarly infrastructures but only provided limited evidence of de facto repository availability \parencite{mannocci_knock_2022}.
Generally, shutting down an information infrastructure should not be considered a failure, because it is ``a highly reflective and introspective set of practices'' \parencite[2205]{steinhardt_breaking_2016} that generates new insights, which can be very useful for future undertakings.
\\
If a repository is integral to the mission of the hosting institution, it is more likely to receive long-term institutional and financial support \parencite{attwood_longevity_2015}. An institution’s mission might often outlast the lifetime of a repository, and, therefore, the institution should make arrangements for the event that a repository has to be closed \parencite{dean_shutting_2016}. If anticipated, the shutdown of a repository can be planned and the stored datasets can be migrated to another repository, while sudden closure might result in permanent data loss \parencite{boyd_understanding_2021}. Therefore, to ensure continuity of access to data, the repository certification organization CoreTrustSeal asks applicants to consider succession planning, the ``preparations for handover of digital objects and services to another repository'' \parencite[13]{coretrustseal_standards_and_certification_board_coretrustseal_2022}.
\\
Migration of data sources is a demanding task, as discrepancies between the legacy and the new system must be overcome \parencite{nyitray_digital_2021}. Although migration is likely common among long-lived research data repositories \parencite{thomer_three_2020}, it is unclear how prevalent the strategy is. The study of biological databases referred to above found that after 18 years, 7 \% of the databases had been migrated to new sites \parencite{attwood_longevity_2015}, but it only covers infrastructures of one discipline. Migration is often initiated in response to changing needs, both of repository operators and users, with the intention to improve repository content, service and management \parencite{thomer_supporting_2018, thomer_maintaining_2022}.

\subsection{Closed repositories in registries}
In a time frame similar to the emergence of repositories, specialized registries collecting information about them have evolved. Presently, there are several of these services that differ in the types of repositories they cover, the metadata schemas they use to describe them, as well as the stakeholders and use cases they serve. The records registries maintain can facilitate repository use, but also monitoring activities. Because repository registries collect information on information infrastructures, they can also be valuable resources for research.
\\
Depending on their mission, registries approach repository shutdowns differently. For example, the registry OpenDOAR focuses on active Open Access repositories. The inclusion criteria of the service state that a repository “must be currently and reliably accessible to any web user around the world” to be indexed.\footnote{OpenDOAR inclusion criteria: \url{https://v2.sherpa.ac.uk/opendoar/about.html}} If editors become aware of a repository shut down, the record of that repository can be made invisible to the public.
\\
In contrast, the registry FAIRsharing, originally focused on databases in the life sciences, does keep record of repositories that have been shut down, since one objective of the service is to monitor the evolution of the objects they describe \parencite{sansone_fairsharing_2019}. The metadata schema therefore reflects the status of repositories, which can be one of \emph{ready}, \emph{in development}, \emph{uncertain} or \emph{deprecated}. If a repository is deprecated, a date and a reason can be specified.\footnote{FAIRsharing record schema: \url{https://fairsharing.github.io/JSONschema-documenter/?schema_url=https://api.fairsharing.org/model/fairsharing_record_schema.json}}
\\
The registry re3data also aims to describe the developments of the global landscape of research data repositories, including the emergence and shutdown of these infrastructures \parencite{pampel_making_2013}. As a result, re3data documents the trajectory of the global repository landscape over time by representing information on the lifespan of a repository via the elements startDate and endDate in the re3data Metadata Schema \parencite[10]{strecker_metadata_2023}:
\begin{itemize}
    \item \emph{startDate} The date the research data repository was released
    \item \emph{endDate} The date the research data repository ended its service of ingesting new research data and/or providing it.
\end{itemize}

\section{Methods}

\subsection{Extracting repository descriptions from re3data}
Due to their overlapping missions, there are considerable intersections between collections of repository registries \parencite{baglioni_semiautomated_2023}. Data collection for this analysis is based on the registry of research data repositories re3data\footnote{re3data: \url{https://doi.org/10.17616/R3D}}, because it is currently the most comprehensive source of information on repositories with a clear focus on research data. Currently, it covers more than 3.000 repositories of all types and from all disciplines. The service was launched in 2012 and therefore has recorded information on the repository landscape for more than 10 years. Repositories are indexed by an international editorial team based on a comprehensive metadata schema. re3data metadata is available under an open license (Creative Commons CC0 1.0 Universal\footnote{Creative Commons CC0 1.0 Universal: \url{https://creativecommons.org/publicdomain/zero/1.0/}}) through an open API.

\subsection{Defining inclusion criteria}
It can be difficult to determine whether a repository is permanently closed – for example, a repository might reopen after being shut down for maintenance, or the repository homepage might have been moved to a new URL \parencite{attwood_longevity_2015}. In addition, research data repositories have multiple temporal properties, such as a life span and temporal coverage, and they can be difficult to differentiate when attempting to determine what constitutes a closed repository.
\\
In this paper, a repository is considered shut down if data is no longer accessible under the original or a new URL, or if the repository website clearly states that the service has ceased operations (while sometimes maintaining very limited access to the data). This definition refers to the life span of a repository and does not relate to the temporal coverage of its collection, unless the website explicitly states that the data are deprecated.
\\
It is important to recognize the ambiguity of the term \emph{closed} in the context of research data repositories. In this paper, \emph{closed} does not refer to restrictions placed on access to data or repository services, but only to the status and life span of repositories.

\subsection{Collecting data}
\begin{figure}[H]
    \centering
    \includegraphics[width=0.5\linewidth]{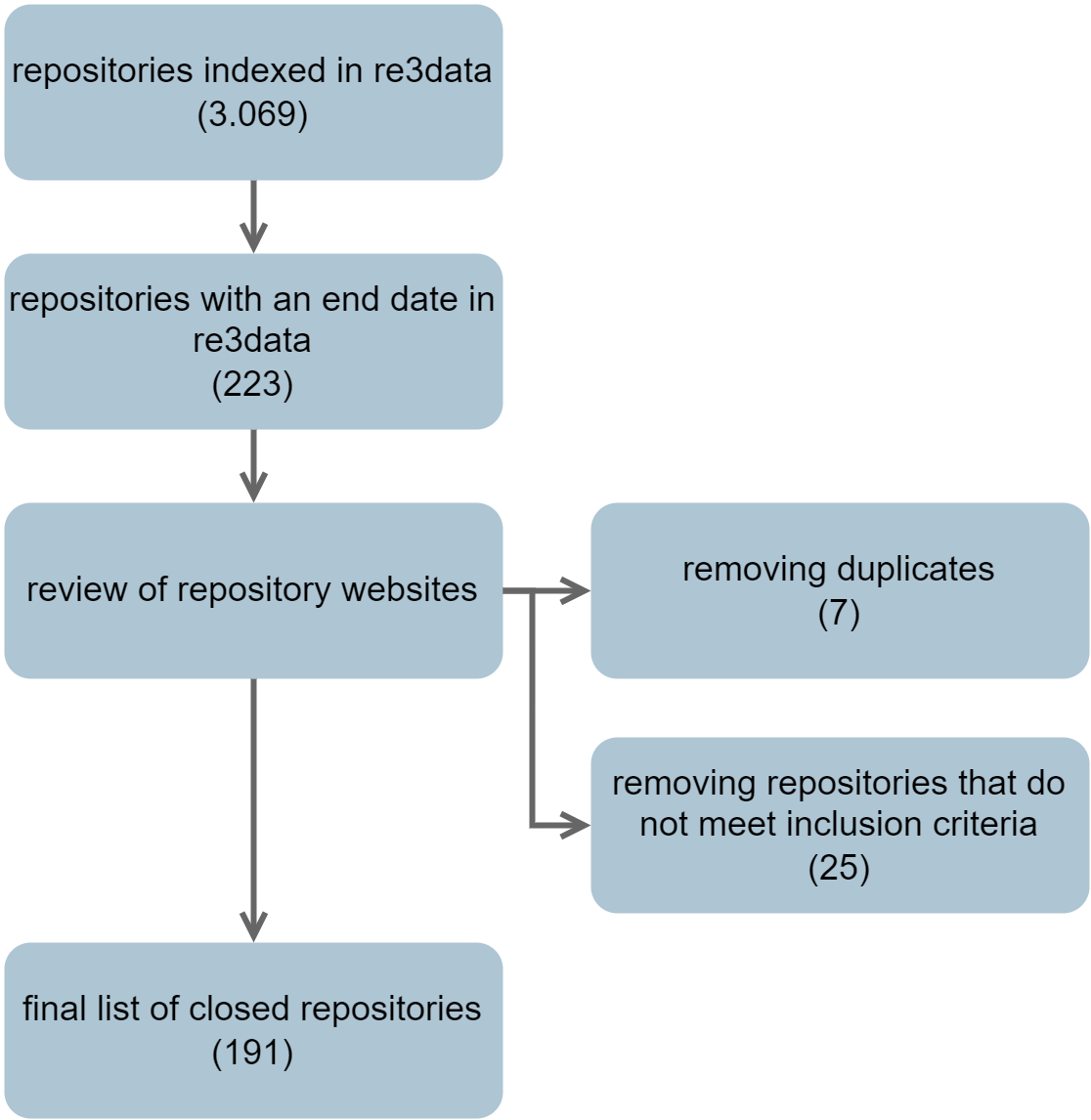}
    \caption{Process of identifying closed repositories}
    \label{fig:collecting_data}
\end{figure}

The analysis draws on re3data repository descriptions and supplements it with information collected from repository websites.
\\
To identify candidates for closed repositories, information on the end date of the 3.069 repositories indexed in re3data was retrieved via the API on 2023, January 2. As shown in Figure \ref{fig:collecting_data}, this list was then restricted to repositories with an end date; this produced a list of 223 candidate repositories. The list of candidates was reviewed by the authors. After visiting the website of each repository on the list, duplicate entries (7) were removed, as well as repositories that did not meet the inclusion criteria for defining closed repositories given above (25).
\noindent
For the remaining 191 repositories, information on the start and end date (year) was retrieved from re3data on 2023, January 2. Given that these properties are optional in the re3data Metadata Schema, the values were verified intellectually and completed where possible.
Between 2023, January 2 and 30, repository websites, both the current version and versions archived by the Internet Archive, as well as additional resources such as data papers describing the repositories were searched for information on the shutdown process. The content analysis of these materials focused specifically on the reason for shutting down and information on the repository taking over custody of the data. This information was collected for each repository in the sample, and, in addition, the availability of data on the current repository websites was checked.
The statements providing reasons for shutting down the repository were generalized and summarized into categories based on the typology of risks to preservation systems developed by Barateiro et al. in Table \ref{tab:risk-typology} \parencite{barateiro_designing_2010}. A similar approach was used previously to study risks to research data in laboratory settings \parencite{kowalczyk_before_2015}.
\\
On 2023, January 30, the type and subject of the closed repositories was retrieved via the re3data API to contextualize the findings \parencite[10--11]{strecker_metadata_2023}:
\begin{itemize}
    \item \emph{type} The type of the research data repository. (for example: disciplinary, institutional)
    \item \emph{subject} The disciplinary focus of the research data repository. (based on subject areas defined by the German Research Organization DFG\footnote{Subject areas defined by the DFG: \url{https://www.dfg.de/en/dfg_profile/statutory_bodies/review_boards/subject_areas/}})
\end{itemize}
The final dataset is openly available \parencite{Strecker_data_2023}.

\section{Results}

\subsection{Prevalence of the repository shut down and time series analysis}
\begin{figure}[H]
    \centering
    \includegraphics[width=1\linewidth]{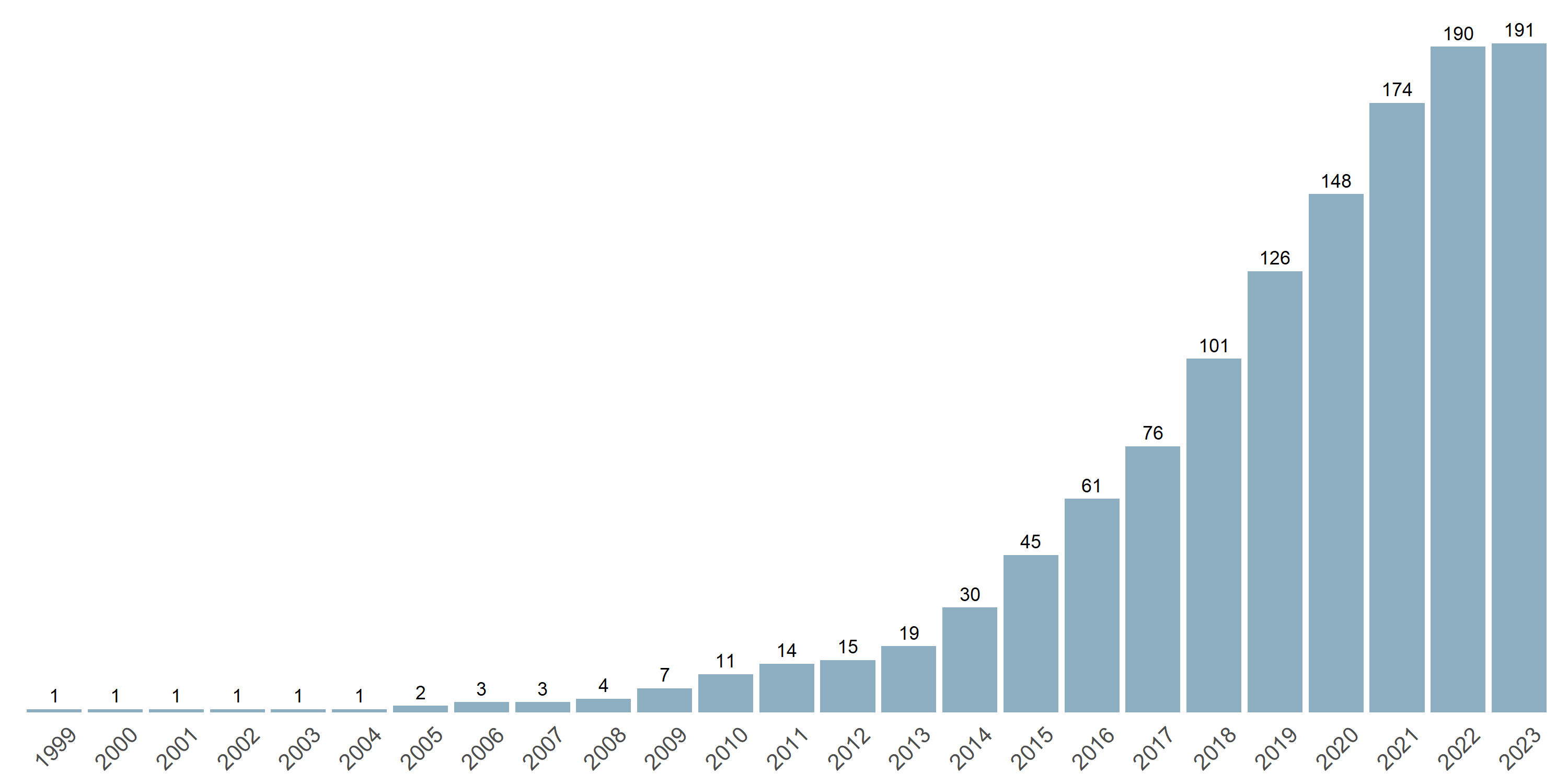}
    \caption{Number of closed repositories indexed in re3data (cumulative)}
    \label{fig:end_date}
\end{figure}

An analysis of the end dates in the sample indicates that re3data has recorded 191 repositories that were closed, with the years of shutdown spanning a period of 25 years. At the time of data collection, this constitutes 6.2 \% of all repositories indexed in re3data (3.069).
\\
The first repository in the sample was shut down in 1999, the next closure occurred after a pause of several years in 2005 (see Figure \ref{fig:end_date}). This is probably linked to the history of re3data -- the service officially took up operations in 2012, and coverage of repositories being shut down before that date is likely sparse. The growth in the number of closed repositories has been fairly consistent from 2012 onward. As of the end of the data collection phase, one repository was closed in 2023.
\noindent
Both a start and end date could be identified for 158 closed research data repositories. Those repositories had been operational between 1 and 57 years before being shut down. The median age of a repository when shut down was 12 years.

\subsection{Characteristics of closed repositories}
\begin{figure}[H]
    \centering
    \includegraphics[width=1\linewidth]{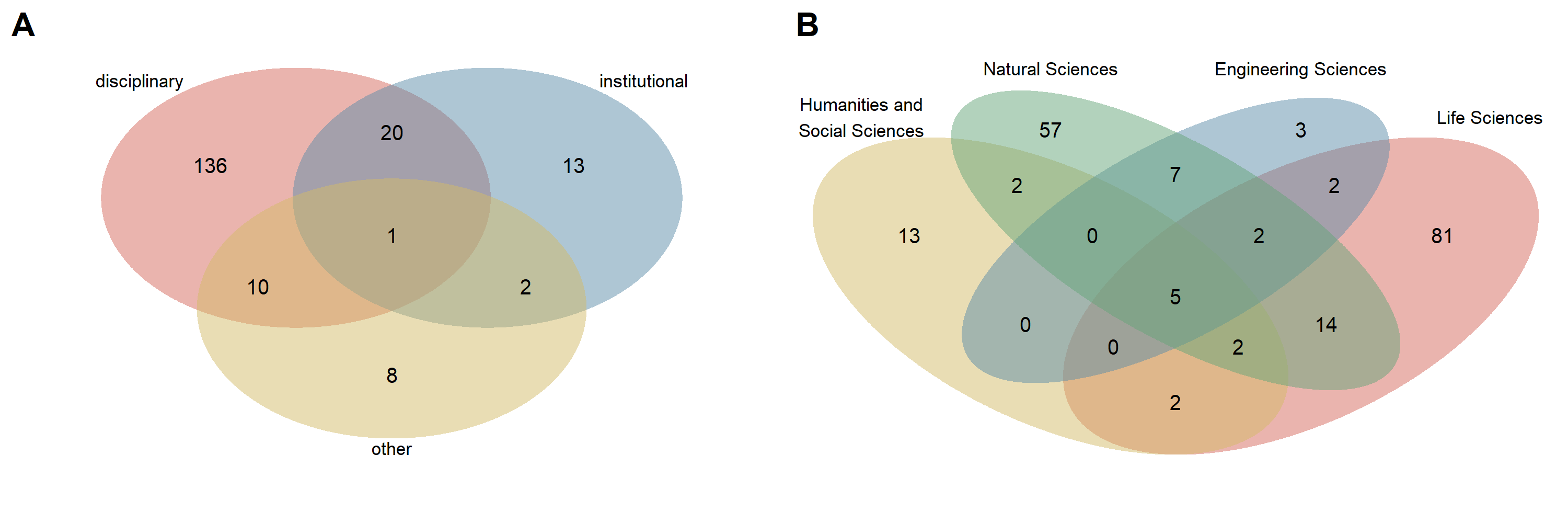}
    \caption{Types (A) and subjects (B) of closed repositories in re3data}
    \label{fig:types_and_subjects}
\end{figure}

As Figure \ref{fig:types_and_subjects} shows, most repositories that were shut down are disciplinary and specialize in data from the life sciences and natural sciences. Compared to all repositories indexed in re3data at the time of data collection, repositories with these characteristics are also overrepresented in the sample. It is important to note that both properties (\emph{type} and \emph{subject}) can be repeated in re3data, so a repository might be assigned more than one type or subject; these combinations are reflected in the Venn diagrams.
\noindent
A comparison of the 25th and 75th percentile of the age distribution of closed repositories shows that compared to long-lived closed repositories, repositories with a shorter life span (25th percentile) were more likely institutional and had a focus on humanities and social sciences (see Table \ref{tab:percentiles}). In contrast, long-lived closed repositories (75th percentile) were more likely disciplinary with a focus on life sciences.
\begin{table}[H]
    \centering
    \begin{tabular}{|l|l|r|r|}
    \hline
        \multirow{4}{*}{type} &  & 25th percentile & 75th percentile \\ \cline{2-4}
                          & disciplinary & 81.6 \% & 95 \% \\ \cline{2-4}
                          & institutional & 23.7 \% & 7.5 \% \\ \cline{2-4}
                          & other & 13.2 \% & 12.5 \% \\ \hline
    \hline
        \multirow{5}{*}{subject} &  & 25th percentile  & 75th percentile \\ \cline{2-4}
                          & humanities and social sciences & 26.3 \% & 5 \% \\ \cline{2-4}
                          & life sciences & 47.4 \% & 57.5 \% \\ \cline{2-4}
                          & natural sciences & 47.4 \% & 47.5 \% \\ \cline{2-4}
                          & engineering sciences & 10.5 \% & 7.5 \% \\ \hline
    \end{tabular}
    \caption{Proportion of type and subject of closed repositories in the 25th (short-lived) and 75th (long-lived) percentile of the age distribution}
    \label{tab:percentiles}
\end{table}

\subsection{Risks resulting in repository shutdown}
The reasons given for the closure of the repository were matched to the typology of risks to preservation systems by Barateiro et al. \parencite{barateiro_designing_2010}.
As Table \ref{tab:reasons} shows, for the majority of closed repositories in the sample (62.5 \% ; 120), a reason for the shutdown could not be determined.
\\
For the remaining 71 repositories, 77 risks could be identified.  Among them, threats (66) were more likely than vulnerabilities (11) to result in shut down, meaning that for these repositories, shutdown can be attributed to specific events rather than weaknesses in the environment.
The most common risks that led to shutdown were managerial threats in nature (64), either due to organizational failure (37) or economic failure (27). Examples for organizational failure include repository shutdown as part of broader reorganization initiatives within the operating organization, or because the mission of the repository was considered fulfilled. Economic failures cover all types of funding cuts, including the cessation of project-related funding. Vulnerabilities of repository technology lead to 5 repositories closing; because repository websites and related materials provided no additional information, both hardware and software obsolescence were identified as risks in these cases.
Threats introduced by external attacks, for example hacking incidents, resulted in repository shutdown in two cases. In one case, vulnerabilities due to media obsolescence lead to repository shutdown -- the research data were considered obsolete and were no longer maintained. Reasons rooted in both vulnerabilities of technology and threats of economic failure were cited by one repository.
\begin{table}[H]
    \centering
    \begin{tabular}{|p{4.5cm}|p{7.5cm}|p{2.5cm}|}
        \hline
        risk & description & number of repositories \\ \hline
        NA & no information available & 120 \\ \hline
        threats - management - organizational failures & repository was shut down as part of broader reorganization initiative within the operating organization, or because its mission is considered fulfilled & 37 \\ \hline
        threats - management - economic failures & repository was closed because funding was cut & 27 \\ \hline
        vulnerabilities - infrastructure - hardware obsolescence & repository was closed because of technological difficulties & 5 \\ \hline
        vulnerabilities - process - software obsolescence & repository was closed because of technological difficulties & 5 \\ \hline
        threats - attacks - external attacks & repository was closed because of acute hacking or security incidents & 2 \\ \hline
        vulnerabilities - data - media obsolescence & repository was closed because the data is considered obsolete & 1 \\ \hline
    \end{tabular}
    \caption{Risks resulting in repository shutdown (more than one reason possible)}
    \label{tab:reasons}
\end{table}

\subsection{Risk of data loss}
The analysis revealed that for 88 \% (168) of the repositories in the sample, data was no longer available on the repository website. 23 (12 \%) repositories still maintained access to data in a limited capacity, for example via a simple FTP interface. Maintenance of data and access services beyond that, including search interfaces, had been ceased. 44 \% (84) of the repositories have listed a repository that has taken over custody of the data. Data loss might have occurred at the closure of 90 (47.1 \%) repositories; these repositories did not maintain (limited) access to data or name a repository that had taken over custody of the data.
\\
One of these cases with a high risk of data loss concerns the repository BIIACS\footnote{BIIACS: \url{https://doi.org/10.17616/R3ZG6K}}. The repository had acquired certification from Data Seal of Approval (DSA), a now discontinued predecessor of the repository certification organization CoreTrustSeal. The repository was launched in 2008, certified by DSA in 2013, and shut down in 2018. It is unknown why the repository was shut down. In the self-assessment document submitted to DSA, the repository expressed “a commitment to maintain the perpetuity of the data with the highest standards, as part of [its] mission”\parencite{leeuw_data_2019}. However, as of today, the data are no longer accessible from the repository website and no repository taking over custody of the data was named. The handles the repository issued to ensure persistent identification of content no longer resolve today.\footnote{Example of a handle that no longer resolves: \url{http://hdl.handle.net/10089/17040}}

\subsection{Data migration}
\begin{figure}[H]
    \centering
    \includegraphics[width=0.7\linewidth]{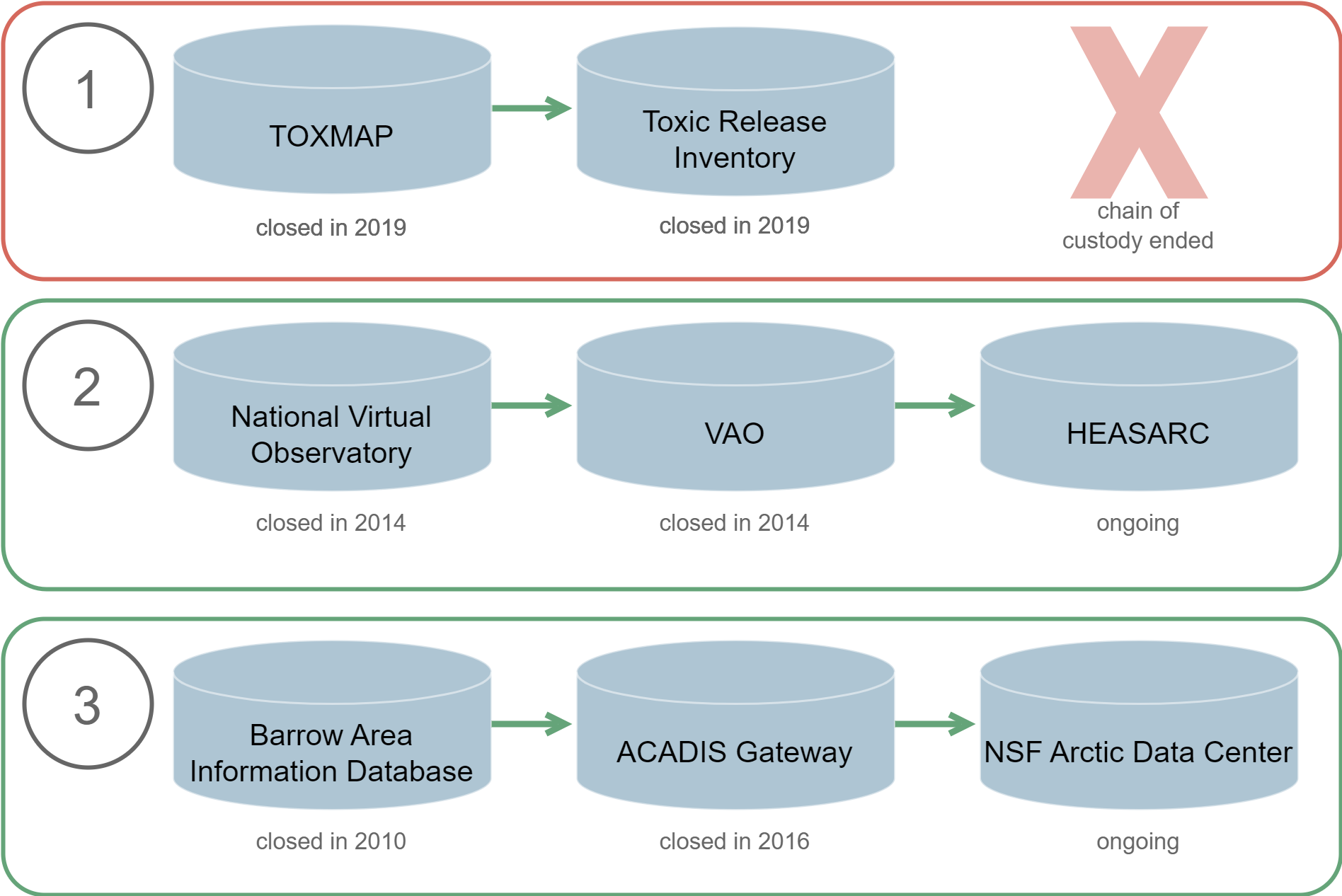}
    \caption{Cases where data was migrated to a new repository that was later shut down; the chain of custody ends in case 1 and continues in cases 2 and 3}
    \label{fig:chains}
\end{figure}

For 84  (44 \%) closed repositories, evidence of data migration was found. In the sample, 75 repositories are listed that have taken over custody of data. Most (60) of these substituting repositories were listed only once, but some were mentioned two (10), three (3) and four (2) times. Three of the repositories mentioned most frequently (PubChem\footnote{PubChem: \url{https://doi.org/10.17616/R3KG65}} (4), Gene Expression Omnibus\footnote{Gene Expression Omnibus: \url{https://doi.org/10.17616/R33P44}} (3), and ArrayExpress\footnote{ArrayExpress: \url{https://doi.org/10.17616/R3302G}} (3)) are large, disciplinary repositories that were established around 20 years ago and are well-known by researchers within the discipline.
\\
8 instances of data migrations reflected in the sample are part of a large-scale reorganization of toxicology data providers within the US National Library of Medicine (NLM) in 2019, where the content of toxicology resources was integrated into other NLM sources \parencite{bolton_toxicological_2020}. Information on chemicals was migrated to PubChem, which affected four repositories in the sample. Data migrations were announced beforehand and are well documented \parencite{kim_pubchem_2021}. A database transition page is still maintained today.\footnote{NLM toxicology database transition page: \url{https://www.nlm.nih.gov/toxnet/index.html}}
\\
The analysis revealed that there are three cases in the sample where custody of data was transferred to a repository that was later shut down (see Figure \ref{fig:chains}). In case 1, the chain of custody potentially ended with the new repository closing, because no evidence was found that the data was migrated again. In the other two cases, data was migrated again to repositories that are still operational today, resulting in a chain of transfers of data custody that is still intact. Figure \ref{fig:chains} shows that in some cases, after a repository was shut down and the data migrated to a new repository, that repository proceeded to be closed in the same year.

\section{Discussion}

\subsection{Life span of research data repositories}
Overall, the emerging landscape of research data repositories is dynamic, with new repositories opening and others being shut down. The analysis showed that research data repositories are shut down fairly frequently: 6.2 \% of the repositories indexed in re3data have been closed. Since re3data took up operations in 2012, repository closures were recorded each year. Consequently, a repository shutdown is not a rare event. As suggested in the literature, shutdown should be considered an expected part of the repository life cycle \parencite{dean_shutting_2016}. It is normal for conditions of data movements to change \parencite{bates_data_2016}, and repository shutdown does not have to be a negative outcome. It can be an appropriate measure that might even result in new insights for the future \parencite{steinhardt_breaking_2016}.
\\
In the sample, the median age of a repository when closing is 12 years. This limited life span could potentially put research data at risk and should be considered when defining retention periods for the long-term preservation of research data. Because repository shutdown is a real threat, infrastructure maintenance should also be reflected in models of digital preservation, and shutdown scenarios planned in advance.
\\
Some closed repositories had particularly long or short life spans. A comparison of top and bottom percentiles of the age distribution revealed differences in type and subject specialization of these long- and short-lived repositories. The sample of closed repositories indicates that disciplinary repositories catering to the life sciences might be more successful at staying operational long-term. However, there are also active long-lived repositories in other disciplines; for example data archives in the social sciences that have been established in the 1960s and are still operational today \parencite{aspray_talking_2019}. Therefore, more research is needed to fully understand this relationship, particularly studies that also consider active repositories with long life spans.
\\
The risks that result in the shutdown of repositories are varied. Some might be anticipated and planned for in advance, such as organizational changes within broader reorganization initiatives, whereas others might not, such as threats from acute security incidents.
\\
As described above, infrastructures operate at different time scales at once by serving current needs and practices while also providing reliable long-term services. It is not clear from the analysis whether long- or short-term requirements put repositories at risk of shutdown more frequently. For example, managerial decisions to reorganize resources within an institution might be motivated by efforts to provide more comprehensive services for current users, or by an attempt to conserve resources in the long term. Overall, more research is needed to determine factors that put research data repositories at risk of being shut down, and how they can be addressed.
However, for most repositories in the sample, the risks resulting in shutdown are unknown. The issue of missing information on repository shutdowns is discussed in more detail below.

\subsection{Identifying closed repositories}
Reviewing repository websites during data collection has demonstrated that it can be difficult to determine whether a repository is permanently shut down. In part, this is due to the fact that a repository has multiple temporal properties, such as a life span and temporal coverage of its collection. These properties are distinct but can be difficult to separate in some cases. For example, a repository might have stopped ingesting new data at a certain point in time, but still maintain access to its collection. Depending on the context, the data might then be considered deprecated, or it might have retained its analytical potential, for example as historical data for time series analysis. Future research could further investigate the interrelation between temporal coverage and life span of research data repositories.
\\
Another factor complicating the identification of closed repositories is the lack of information on planned downtime. A repository might be temporarily unavailable, for example due to maintenance activities. In case of prolonged downtime periods where no information on the maintenance schedule is given on the repository website, repository users might assume the repository is permanently closed and report this to registries. In the sample, there was at least one repository that had an end date in re3data, but came back online after an unannounced downtime period.\footnote{TreeBASE: \url{https://doi.org/10.17616/R3DK58}} Repositories can avoid this by announcing planned downtime on their websites.

\subsection{Strategies for preventing data loss}
The analysis of information collected from repository websites focused on two strategies for preventing data loss when a repository is shut down: Maintaining limited access to data and migrating data to another repository. The results show that these strategies are used by repositories to varying degrees.
\\
Most repositories do not uphold access to data when they are shut down, often leaving no trace of the repositories or their contents on their websites. Only few repositories opt to maintain limited access to data, for example via a simple FTP interface. This strategy, however, is not a permanent solution, since all curation and long-term preservation activities have been discontinued. If no measures are taken to preserve individual datasets, it becomes more likely as time passes that data will no longer be usable. In addition, browser support for FTP is declining, which might limit human access to data in the future.
\\
In contrast, the strategy of migrating data to another repository is more common, being used by almost half of the repositories in the sample. Overall, this strategy is more likely to retain the usefulness of data, because the repositories that have taken over custody of the data can apply appropriate curation and preservation measures. However, this strategy should not be considered definitive either, as the burden of infrastructure maintenance is not eliminated but transferred to the succeeding repository. This is demonstrated by three cases of chaining of data migration, meaning that custody of data has been transferred several times. In two of these cases, the chain of data custody remained intact, but in one case it ended because the repository that had ingested the data was shut down without indicating a successor.
\\
Some repositories seem to have established themselves as reliable options for taking over custody of data, given that they have ingested data from multiple repositories that were shut down. These tend to be large repositories with a disciplinary focus and a comparatively long life span. They seem to have become central infrastructures for the research domain they focus on while also serving as a safe haven for data collections from other repositories. Future research should critically investigate the long-term consequences of data migration. For example, data migration conducted within larger reorganization efforts can affect a number of repositories and result in a consolidation of the repository landscape, by combining collections into large, central repositories. Central repositories might be able to use limited resources more efficiently, but could also create single points of failure if they come at risk of being shut down themselves.
\\
Other measures also have the potential to reduce the risk of data loss, for example repository certification. One repository in the sample acquired formal certification from DSA, but was shut down despite the stated commitment to ensuring long-term availability of data, demonstrating that obtaining certification is no guarantee that a repository will stay operational long-term. However, the certification process can require repositories to provide evidence of sustainable operations and succession planning, encouraging them to implement appropriate measures. The same case also highlights the limitations of another measure to ensure sustained access to data: The use of persistent identifiers. Persistent identifiers are very useful for reliably referring to research data, but they have to be maintained to function as intended.

\subsection{Documenting repository shut downs and changes in data custody}
The analysis demonstrates that most repositories completely disappear after being shut down: They maintain neither limited access to data nor a web page with information on the shutdown process or potential data migration. This lack of information can have serious consequences for data citation and the permanence of the scholarly record. If research data are reused and cited, but later become unavailable due to the repository shutting down, references can break, which erodes the scholarly record. This is especially the case if no information about the repository shutdown or data migration is shared.
\\
An example of how information on repository shutdown can be provisioned is the comprehensive reorganization of toxicology information within the NLM. Even years after the initiative was concluded, the NLM maintains a database transition page, a website that provides information on the reorganization process. The initiative was announced in advance, and data migrations are documented in detail, enabling researchers to trace datasets to the sites they are stored at now.
\\
Registries can contribute to making information on repository shutdown and changes in data custody accessible. As services that collect information on research data repositories and their trajectories, they are uniquely positioned to do so. For example, re3data currently indicates whether a repository was closed, and when. The most recent version of the re3data Metadata Schema, version 4.0, introduces the option to reflect transfers of data custody \parencite{strecker_metadata_2023}, which will make data migration visible in the registry.
\\
To realize this vision of more comprehensive information on repository shutdown, repository operators should consider being more transparent about the process. By providing information on the shut down process beforehand and naming the succeeding repository if the data were migrated, they can increase transparency, help researchers trace chains of data custody, and inform other repository operators.

\section{Conclusion}
The analysis showed that research data repository shutdown is not a rare phenomenon but should be considered an integral part of the life cycle of repositories. Repository shutdown poses a real threat to the perpetual availability of research data. Therefore, when planning preservation measures, repository operators should also take into account the infrastructure perspective on the long-term availability of data. Planning ahead increases the chance of saving data in case the repository is forced to shut down. Strategies like data migration can prevent immediate data loss but should not be considered permanent solutions. To more fully reflect the landscape of research data repositories and support the integrity of the scholarly record, registries can document repositories shutting down as well as changes in data custody. Overall, more research is needed to evaluate strategies for preventing data loss and to understand how specific factors affect the risk of repository shutdown, for example how deeply a repository is embedded in its community, its revenue streams, and its ability to respond to both short- and long-term shifts in user needs.

\section*{Limitations}
Since re3data took up operations in 2012, re3data records likely don’t document all incidents of repositories being shut down before that time.
The analysis is based solely on information that is publicly available online. Most closed repositories have provided only limited or no information about the shutdown process, therefore, data migration events or reasons for shutting down are likely underreported.

\section*{Funding information}
This work has been supported by the German Research Foundation (DFG) under the project \emph{re3data COREF} (Grant ID 422587133).

\newpage
\printbibliography

\end{document}